\documentclass[pra,aps,twocolumn,showpacs,superscriptaddress,floatfix,amsmath,amssymb,nofootinbib]{revtex4}
\usepackage{amsfonts}
\usepackage{graphicx}
\newcommand{\beq}{\begin{equation}}
\newcommand{\eeq}{\end{equation}}

\newcommand{\beqa}{\begin{eqnarray}}
\newcommand{\eeqa}{\end{eqnarray}}

\begin{document} 
\title{Energy consumption for  Shortcuts To Adiabaticity}
\author{E. Torrontegui}
\email{eriktm@iff.csic.es}
\affiliation{Instituto de F\'{\i}sica Fundamental IFF-CSIC, Calle Serrano 113b, 28006 Madrid, Spain}
\affiliation{Institute of Chemistry and The Fritz Haber Research Center, The Hebrew University, Jerusalem 91904, Israel}
\author{I. Lizuain}
\affiliation{Department of Applied Mathematics, University of the Basque Country UPV/EHU, Plaza Europa 1, 20018 Donostia-San Sebastian, Spain}
\author{S. Gonz\'alez-Resines}
\affiliation{Departamento de Qu\'{\i}mica F\'{\i}sica, Universidad del Pa\'{\i}s Vasco - Euskal Herriko Unibertsitatea, 
Apdo. 644, Bilbao, Spain}
\author{A. Tobalina}
\affiliation{Departamento de Qu\'{\i}mica F\'{\i}sica, Universidad del Pa\'{\i}s Vasco - Euskal Herriko Unibertsitatea, 
Apdo. 644, Bilbao, Spain}
\author{A. Ruschhaupt}
\affiliation{Department of Physics, University College Cork, Cork, Ireland}
\author{R. Kosloff}
\affiliation{Institute of Chemistry and The Fritz Haber Research Center, The Hebrew University, Jerusalem 91904, Israel}
\author{J. G. Muga}
\email{jg.muga@ehu.es}
\affiliation{Departamento de Qu\'{\i}mica F\'{\i}sica, Universidad del Pa\'{\i}s Vasco - Euskal Herriko Unibertsitatea, 
Apdo. 644, Bilbao, Spain}
\begin{abstract}
Shortcuts to adiabaticity let a  system reach the results of a slow adiabatic process in a shorter time.
We propose to quantify the ``energy cost'' of the shortcut 
by the energy consumption of the system enlarged by including the  control device. A mechanical model where the dynamics of the system and control device can be explicitly described  illustrates  that a broad range of possible values for the consumption are possible, including  zero (above the adiabatic energy increment) when  
friction is negligible and the energy given away as negative power  is stored and reused by perfect regenerative braking.  
\end{abstract}  	
\maketitle
%
 
%
%
%
%
\begin{figure}[t]
\begin{center}
\includegraphics[width=4.cm]{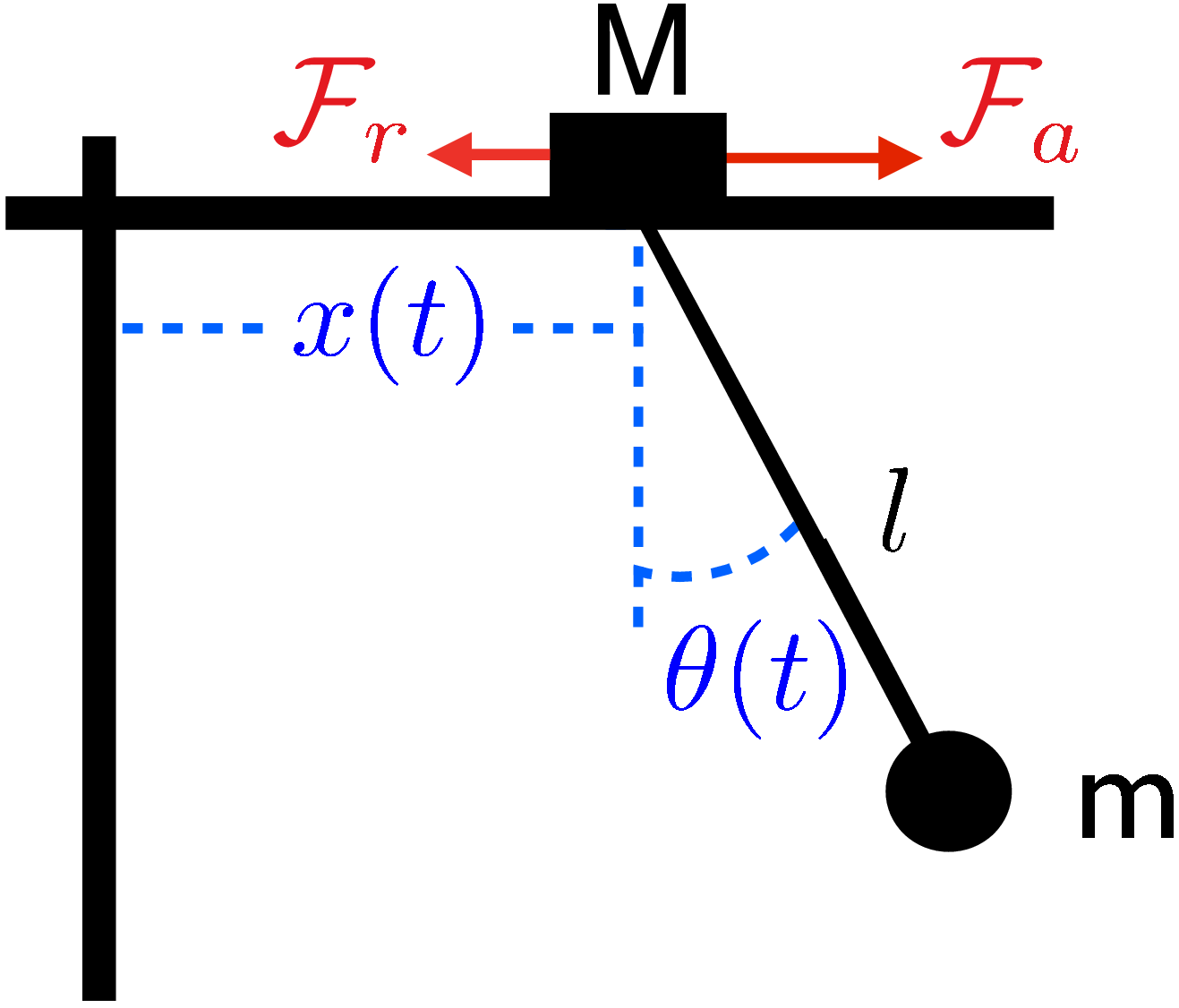}
\caption{(Color online) 
Overhead crane composed by a load of mass $m$  and a trolley of mass $M$ connected through a rope of constant length $l$. The red-solid arrows represent the active force ${\mathcal{F}}_a$ and the friction force ${\mathcal{F}}_r$ acting on 
a rightwards moving trolley.}
\label{crane}
\end{center}
\end{figure}
%
%
%
\section{Introduction}
Shortcuts to adiabaticity (STA) \cite{PRL2010,review} are protocols for the time dependence of the control parameters of a system (hereafter primary system, PS)  
so that it reaches the same final conditions (energy, populations, or state) of a slow adiabatic process
in a shorter time. STA have found widespread applications in atomic, molecular and optical physics and 
beyond, e.g. for classical systems \cite{clas,Kazu,FF_inva}, as a generic tool to combat decoherence and design robust, fast processes or devices. 
{Some STA use the   structure of the Hamiltonian describing the slow process for the PS, as in invariant-based methods \cite {PRL2010}, and others add new control terms, as in counterdiabatic approaches \cite{Rice}, but this distinction does not affect the following  discussion.} 
 
The total mechanical work done on the PS in a given STA is by definition equal to the work done in the adiabatic process, 
i.e., the adiabatic energy increment between initial and final states. 
It was soon  clear that this quantity could not represent all relevant energy flows, 
which lead to consider  alternative measures \cite{CM}. 
Several, disparate  definitions of energy cost have been proposed in the context of quantum thermodynamics to characterize   
quantum engines and refrigerators \cite{adolfo,sarandy1,sarandy2,poletti,lutz,adc,entropy,wiebe,CD,bt,al}. 
These definitions  have been  systematically formulated in terms of the cycling system (PS) alone. 
Even if the existing proposals have
their own merits and applications,  the point of view put forward in this
article is that a broader perspective is necessary for the definition to be useful and practically relevant,  
addressing not only the PS, but the 
control system (CS) that drives the time-dependent parameters. In other words,  
we advocate to redefine and expand the  ``system'' in the model so as to include the PS and the CS
into an enlarged system.
It might appear that this simply shifts the  system-defining border  so that the same problem is translated
towards  the new border. The important point is to find a meaningful divide, for which the energy changes with the outer world  are  modeled by forces that can be easily translated into fuel or electric power consumption by an active device.   
Such a  shift is crucial to make the energy ``cost" a significant quantity that indeed  has something to do with 
the  feasibility of the processes, minimal times allowed, or  economic costs. Some examples help to clarify this: If a train (CS)  transports 
cargo  (PS) horizontally between two stations, the total energy increment of the cargo is zero.  
Surely what interests us more as a relevant  cost is the energy consumption by the active force that the engines should do, translated into fuel consumption. We thus need to  evaluate this force by expanding the physical model to include the train itself, taking into account friction, the braking mechanism, and paying attention 
to the maximum power deliverable by the engine, that will put limits on the minimal transport times.                  
Similar examples can be drawn from studies by nutritionists or biomechanicists 
concerned with the kilocalories the  body consumes, or the oxygen 
intake,  to perform  a given task or exercise \cite{biomech}.   
For a weightlifter (CS) pushing a weight (PS) up,  the energy expenditure does not only depend on the work done on the weight, but on CS-dependent factors such as the lifter skill and weight, and muscular mass.  
          
This paper is based on a simple model for which enlightening, explicit expressions for the dynamics, power, and energy consumption are worked out.  
In Sec. \ref{secII} we present our model, a mechanical crane, and the main results. The model is described by equations similar to the ones 
used for the transport of neutral atoms or ions in microscopic traps. 
In Sec. \ref{minimal} we find the optimal protocol with respect to energy consumption, 
and the paper ends with a discussion where we surmise the implications, that we expect to be broadly applicable.        
\section{Model and results\label{secII}}
The model is an overhead crane, as depicted in Fig. \ref{crane}, composed by a trolley of mass $M$ (CS) moving along a horizontal bridge and a load of mass $m$ (PS) pending by a constant-length rope \cite{xia}. 
We neglect the stiffness and mass of the rope, and air resistance. The load can be regarded, in the small oscillations regime 
characteristic of these devices,  as a harmonic oscillator with moving center.  
The generalized coordinates are the position of the trolley $x(t)$ and the swing angle $\theta(t)$. The process we consider is 
a transport of the load by moving the trolley from $x=0$ to $x=d$ in a time $t_f$. If done adiabatically, the initial and final energies of the load should be equal.   
Shortcuts for quantum systems subjected to a moving harmonic (or otherwise) trap have been extensively studied, see e.g.  
\cite{Masuda,transport,OCTtransport}.                  
 
The external forces depicted in Fig. \ref{crane} are 
the actuating force $\mathcal{F}_a$ (e.g. due to an engine, or to a braking mechanism if it opposes the direction of motion of the trolley) 
and the friction modelled here as $\mathcal{F}_r=-\gamma\dot{x}$, $\gamma\ge 0$. 
The Lagrangian, without friction, is $\mathcal{L}=\mathcal{L}_1+\mathcal{L}_2$, 
%
\beqa
\label{L}
\mathcal{L}_1&=&\frac{m}{2}\bigg[\dot x^2+l^2\dot\theta^2+2l\dot x\dot\theta\cos\theta\bigg]\nonumber+mgl\cos\theta,
\nonumber\\
\mathcal{L}_2&=&\frac{M}{2}\dot x^2+{\mathcal{F}}_a x,
\eeqa
where the dots represent  time derivatives, $l$ is the rope length,  and $g$ the gravitational acceleration. 
With friction, the equations of motion are derived from the Euler-Lagrange equations,
with the equation on the trolley position modified to include a friction term, $\frac{d}{dt}(\frac{\partial L}{\partial \dot{x}})-\frac{\partial L}{\partial x}+\frac{\partial {\cal F}}{\partial {\dot{x}}}=0$, 
where ${\cal F}=\gamma \dot{x}^2/2$ is Rayleigh's dissipation function \cite{Gold}, 
\beqa
0&=&l\ddot\theta+\ddot x\cos\theta+g\sin\theta,
\label{dyn1}
\\
\mathcal{F}_a+\mathcal{F}_r&=&M\ddot x+m(\ddot x+l\ddot\theta\cos\theta-l\dot\theta^2\sin\theta). 
\label{dyn2}
\eeqa
Equation (\ref{dyn1}) defines the kinematics  of the load in terms of  $x(t)$ only,
i.e., it is formally independent of characteristics of the trolley such as mass, or friction, for a given $x(t)$. 
This allows the formal treatment of the load as an open 
system subjected to external time-dependent control, but  $x(t)$ depends on these characteristics,  and on the angle and the pulling force  via  Eq. (\ref{dyn2}).  
We may compute the frictionless Hamiltonian of the total system through the Lagrangian $\mathcal{L}=\mathcal{L}_1+\mathcal{L}_2$ given by Eq. (\ref{L}),
$
\mathcal{H}=\dot xp_x+\dot\theta p_{\theta}-\mathcal{L},
$
%
where $p_x=\partial\mathcal{L}/\partial\dot x$ and $p_{\theta}=\partial\mathcal{L}/\partial\dot \theta$. 
To account for friction, one of Hamilton's equations changes to \cite{mech,friction}
$\dot{p}_x=-\frac{\partial {\mathcal{H}}}{\partial x}-\frac{\partial {\cal F}}{\partial \dot{x}}$. 
The power produced by the force $\mathcal{F}_a$ 
can be expressed as the rate of change of  
$\mathcal{H}_0=\mathcal{H}+\mathcal{F}_ax$ (the last term cancels the external interaction
$-\mathcal{F}_a x$ in $\mathcal{H}$ to leave the bare mechanical energy),
 plus the energy loss rate due to friction, 
\beq
\label{P2}
\mathcal{P}=\frac{d{\cal{H}}_0}{dt}+\gamma \dot{x}^2=\mathcal{F}_a\dot x. 
\eeq
The total derivative is computed along the trajectory making use of  Hamilton's equations for ${\mathcal{H}}$ modified by the friction term. 
Here a meaningful divide is established, with the relevant connection to the outer world being a force $\mathcal{F}_a$ produced by an external engine that, for positive power, consumes fuel to increase the internal mechanical energy and fight against friction.  
The total energy consumption
could  be defined 
as  the integral of the power \cite{xia}, but  this would ignore the peculiarities of braking phases where ${\mathcal{F}}_a$ and $\dot{x}$ have different signs. 
We propose instead a more realistic expression parameterized by $-1\le\eta\le1$, 
that depends on the braking mechanism, 
\beq
\mathcal{E}=\int_0^{t_f}\!\! dt \mathcal{P_+}+\eta \int_0^{t_f}\!\! dt  \mathcal{P_-}={\mathcal{E}}_++\eta{\mathcal{E}}_-,
\label{Econsum}
\eeq
where $\mathcal{P_\pm}=\Theta(\pm\mathcal{P})\mathcal{P}$  are positive/negative parts of $\mathcal{P}$ for  
accelerating  or braking phases of the trolley motion, and $\Theta$ is the Heaviside function. ${\mathcal{E}}_\pm$ are positive and negative parts of the integral.   
While more sophisticated descriptions are possible, with 
$\eta$ depending on several variables, our aim here is to set a crude model that captures 
the essence of the energy trade during braking, and provides limiting scenarios:  
$\eta=1$ corresponds to a mechanism  able to fully accumulate the braking energy ${\mathcal{E}}_-$ and give it back on demand, i.e., perfect regenerative breaking; 
$\eta=-1$ corresponds to using the engine in both phases of the motion, whereas $\eta=0$ is the limit in which braking fully dissipates
the energy loss of the system with negligible energy consumption.  



To find STA we use the horizontal deviation of the load from
the trolley position, $q(t)=l\sin\theta(t)$, and assume the small oscillations regime.  Equation (\ref{dyn1}) becomes
\beq\label{gen}
\ddot q+\omega^2q=-\ddot x,
\eeq
%
where $\omega^2=g/l$. 
The dynamics of  the load (PS) is described in a moving frame by a forced harmonic oscillator, 
which can be derived from the Hamiltonian
\beq
\label{ham}
H=\frac{p^2}{2m}+\frac{1}{2}m\omega^2q^2+m\ddot xq,
\eeq
where $p=m\dot q$ is the canonical momentum of $q$. Associated with $H$ there is the invariant of motion \cite{LL} 
\beq
\label{inv}
I=\frac{1}{2m}(p-m\dot\alpha)^2+\frac{m}{2}\omega^2(q-\alpha)^2,
\eeq
where 
$\alpha(t)$ is an auxiliary  trajectory that must follow the dynamics of a forced harmonic oscillator \cite{LL},
\beq
\label{qho}
\ddot\alpha+\omega^2\alpha=-\ddot x.
\eeq
%
%
%
%
%
\begin{figure}[t]
\begin{center}
\includegraphics[width=6cm]{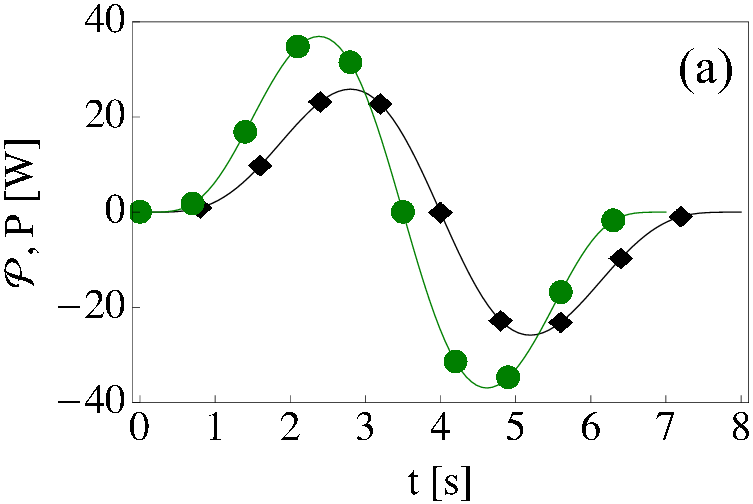}
\includegraphics[width=6cm]{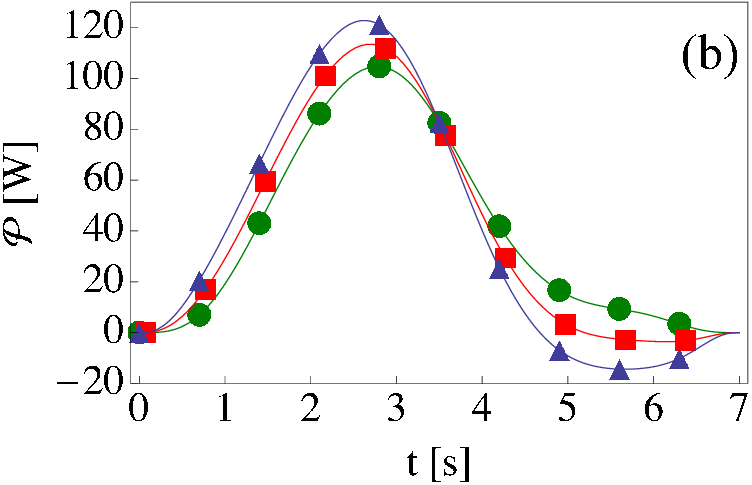}
\caption{(Color online) The total power $\mathcal{P}$  to control $x(t)$ for different $M$  and 
friction coefficients, and load power $P$.
Symbols represent $\mathcal{P}$ using  the small-oscillation approximation, Eq. (\ref{Preal}), and lines using the exact Eq. (\ref{P2}). 
(a) Power $P$ of the load (symbols) and total power $\mathcal{P}$ (lines) in the $M=\gamma=0$ limit for $t_f=7$ s (green line and circles) and $t_f=$ 8 s (black line and diamonds). 
(b) Total power $\mathcal{P}$ with friction, $\gamma=15$ kg/s, $t_f=7$ s, for different values of the trolley mass: 
$M=0$ kg (green-solid line and circles),  $M=10$ kg (red-solid line and squares), and  $M=20$ kg (blue-solid line and triangles). 
$m=10$ kg, $l=5$ m, $d=10$ m, 
$q(0)=0$ m, $\dot q(0)=0$ m/s, and $g=9.8$ $m/s^2$.}
\label{power}
\end{center}
\end{figure}
%
%
%
%
%
We  choose $\alpha(t)$ functions that satisfy the boundary conditions (b. c.) $\alpha(t_b)=\dot\alpha(t_b)=\ddot\alpha(t_b)=0$, for $t_b=0,t_f$. 
In this way
$\ddot x(t_b)=0$ and, from Eqs. (\ref{ham}) and (\ref{inv}), $H(0)=I(0)=E_0$ for any arbitrary trajectory $q(t)$ satisfying Eq. (\ref{gen})
with  initial energy $E_0$ 
(for the auxiliary trajectory $\alpha$, $E_0=0$). As $I$ is invariant,  $I(t_f)=E_0$. Moreover the final energy is $H(t_f)=I(t_f)=E_f$. In summary, imposing the appropriate b. c. on $\alpha$,  $E_f=E_0$ for any trajectory, as for an adiabatic,  slow process, but in a finite time. 

We interpolate $\alpha(t)$ with a polynomial, $\alpha(t)=\sum_{i=0}^7a_it^i$, where the first six coefficients ($a_0-a_5$) are derived from the six  b. c. for  $\alpha$. The trajectory $x(t)$ of the trolley is deduced from Eq. (\ref{qho}),
%
$x(t)=-\int_0^t\!\!dt'\int_0^{t'} \!\! dt''[\ddot\alpha(t'')+\omega^2\alpha(t'')]$,
%
and satisfies $\ddot x(t_b)=\dot x(0)=x(0)=0$. The   coefficients $a_6$ and $a_7$ are set by demanding $\dot x(t_f)=0$ and $x(t_f)=d$. Due to the freedom to design  $\alpha$, 
optimal control theory could be used to find trolley trajectories that optimize  
a chosen variable given some physical constraints \cite{OCTtransport}.
%
%
\begin{figure}[t]
\begin{center}
\includegraphics[width=6cm]{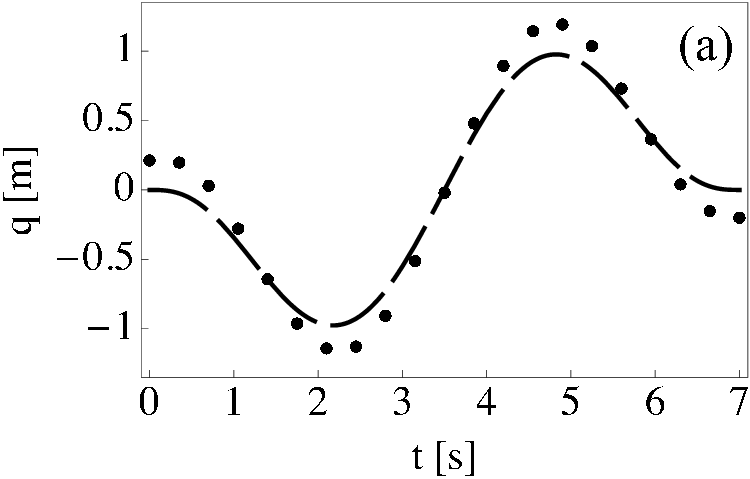}
\includegraphics[width=6cm]{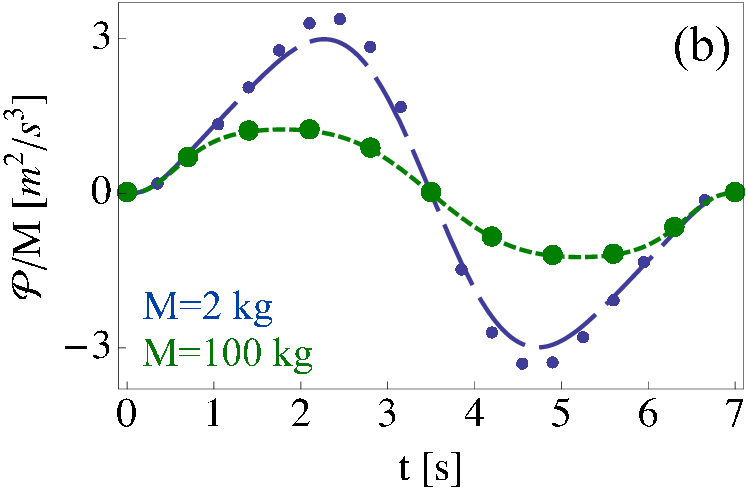}
\caption{(Color online)  Effect of the trajectory $q(t)$ on the total power $\mathcal{P}$ for
different trolley masses. $q(0)=\dot q(0)=0$ (dashed-lines), $q(0)=0.2$ m, $\dot q(0)=0.1$ m/s (circles).
(a) $q(t)$ for different initial conditions. 
(b) Corresponding power consumed for different $m/M$ ratios: $M=2$ kg (blue long-dashed line and small circles); 
$M=100$ kg (green short-dashed line and big circles). 
$m=1$ kg, $l=5$ m, $d=10$ m, $t_f=7$ s,
$\gamma=0$ kg/s, and $g=9.8$ $m/s^2$.}
\label{Qindep}
\end{center}
\end{figure}
%
%
For small oscillations, the total power in Eq. (\ref{P2}) takes the form 
\beqa
\label{Preal}
\mathcal{P}&=&\left(M\ddot x-mq \omega^2+\gamma \dot{x}\right)\dot x,
\eeqa
plotted in Fig. \ref{power} for  $q(t)=\alpha(t)$. The terms in parenthesis represent the force to move a free trolley (with no load or friction), minus the force that the load exerts on the trolley (a ``pull or drag'' backaction whose sign depends on their relative positions), minus the friction force (which  always gives a positive contribution to the power).   
Let us compare this quantity to the power on the load,  
%
$
P=\frac{dE(t)}{dt},
$
where $E(t)$ is the mechanical energy of the load, $E(t)=m(\dot{x}+\dot{q})^2/2+m\omega^2q^2/2$ (For arbitrary $t$, this is different from $H(t)$, since $H$ is defined in a moving frame, but they coincide at the boundary times.). Using Eq. (\ref{gen}), $P=-mq\omega^2  \dot{x}$,    
%
%
%
which is the rate of energy change in the PS  
but, for a given $x(t)$,  it ignores other features of  the trolley. 
In contrast, ${\mathcal{P}}$ and  ${\mathcal{E}}$  generally depend, see Eq. (\ref{Preal}), on the characteristics of the CS ($M$,   $\gamma$), on its dynamics  ($\dot{x}$, $\ddot{x}$), and  on the deviation  of the load, $q(t)$. 
If $M=\gamma=0$,
%
$
\mathcal{P}_{M=\gamma=0}=-mq \omega^2 \dot x=P, 
$
%
see Fig. \ref{power}. 
A practical advantage of the limit $M>>m$ is that $\mathcal{P}$ can be made essentially independent
of $q(t)$, i.e., on the initial conditions $\{q(0),\dot{q}(0)\}$, see Fig. \ref{Qindep}, where the $\alpha(t)$ chosen implies that $\ddot{x}=0$ at the boundary times, 
and at the middle time. This stabilization comes with a price, namely, higher power peaks due to a larger $M$.    
%
%

%
%
%
%
\begin{figure}[t]
\begin{center}
\includegraphics[width=4.25cm]{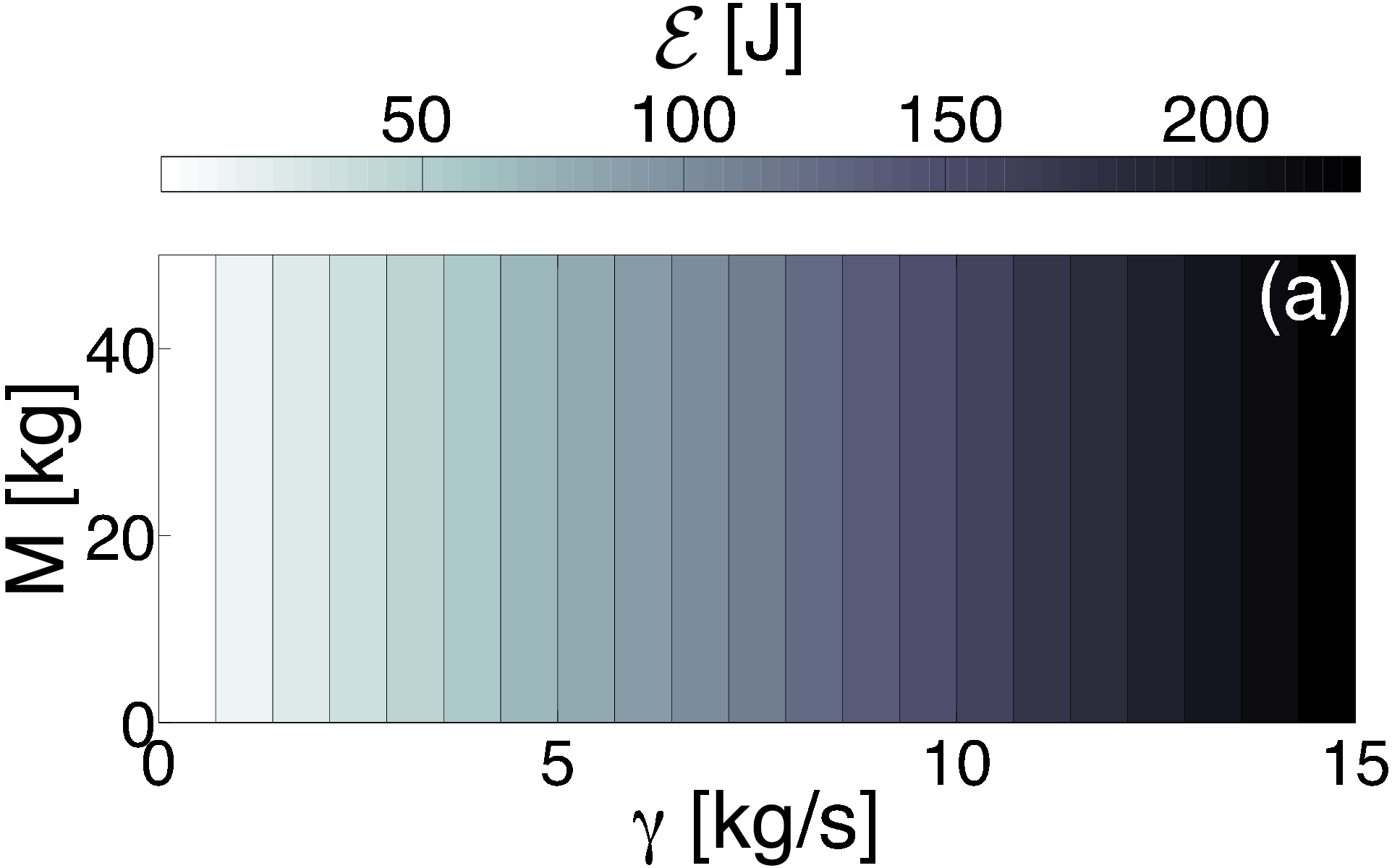}
\includegraphics[width=4.25cm]{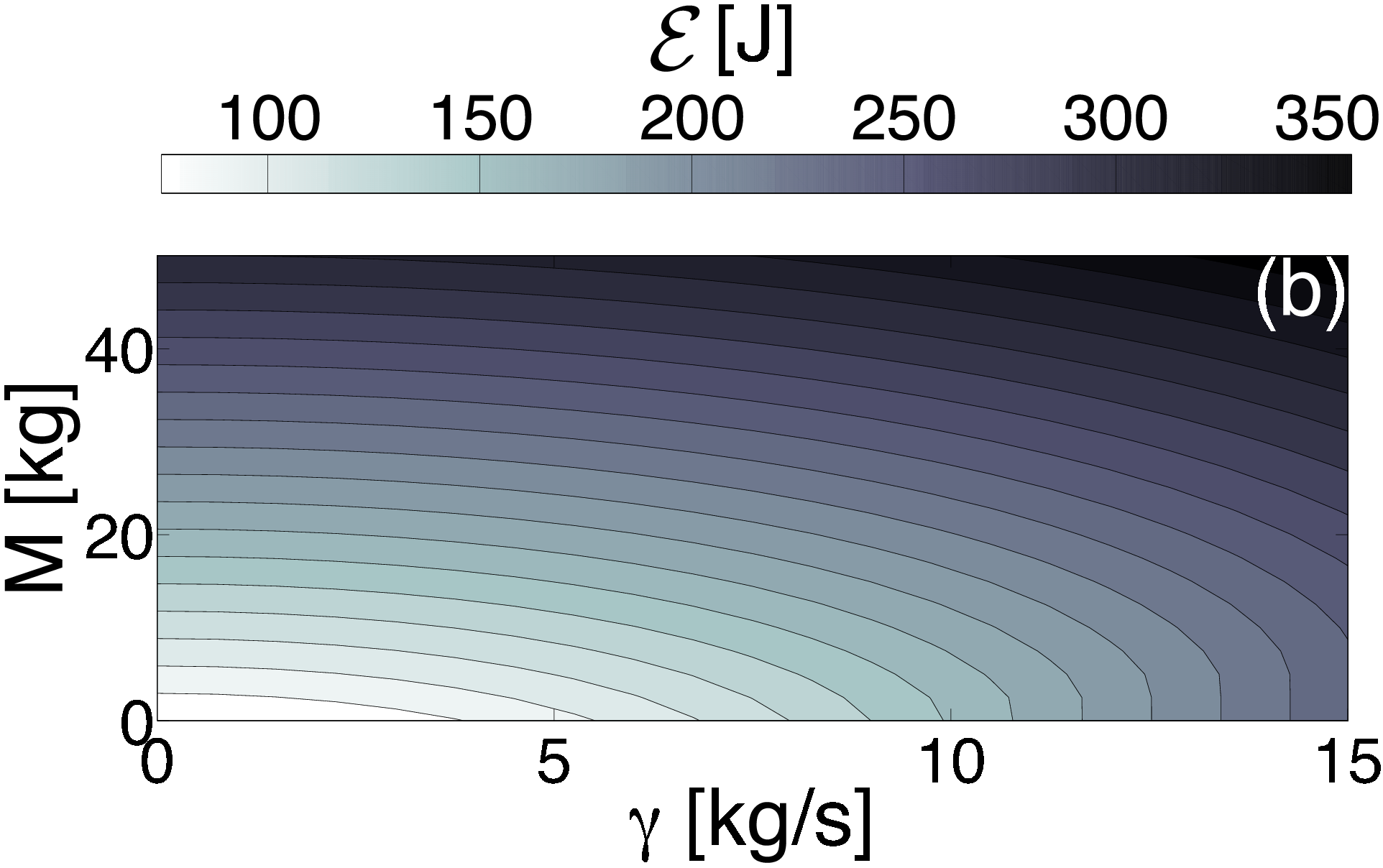}
\caption{(Color online) Contour surface of the energy consumption $\mathcal{E}$ as a function of the CS variables $M$ and $\gamma$ for different values of the $\eta$ parameter.
(a) $\eta=1$ and (b) $\eta=-1$.  
$m=10$ kg, $l=5$ m, $d=10$ m, $t_f=9$ s,
$q(0)=0$ m, $\dot q(0)=0$ m/s, and $g=9.8$ m/s$^2$.}
\label{energy}
\end{center}
\end{figure}


The integral of ${\mathcal{P}}$, without friction, $\gamma=0$, is zero by construction of the STA (the final adiabatic
energy of the load must be equal to the initial one, and the trolley starts and ends at rest),
so the total energy consumption  would be zero for $\eta=1$. The other parameters 
may be arbitrary, even $t_f$, within small oscillations. Friction, and realistic braking mechanisms ($\eta\ne 1$) imply 
$|\eta {\mathcal{E}}_-|<{\mathcal{E}}_+$ and therefore dependences of ${\mathcal{E}}>0$ on $\gamma$, $M$, or $t_f$.  Note that ${\mathcal{E}}$ depends linearly on $\eta$ with minimum ${\mathcal{E}}_++\mathcal{E}_-$ at $\eta=1$, and maximum ${\mathcal{E}}_+-\mathcal{E}_-$ at $\eta=-1$. Since the time integral of the frictionless part of Eq. (\ref{Preal}) is
zero,  we get, using the Euler-Lagrange equation, 
the lower bound 
\beq
\label{lower}
{\mathcal{E}}\ge \gamma d^2/t_f, 
\eeq
valid for all $\eta$. This agrees with Landauer's expectation on energy costs
of processes not involving information losses \cite{Landauer}.   
However, a different, tighter bound is found in Sec. \ref{minimal} from the optimal protocol.     

Some trends are seen  in Figs. \ref{power} and \ref{energy}:
friction enhances ${\mathcal{E}}_+$ and diminishes, or even suppresses ${\mathcal{E}}_-$;  a larger  $M$ generally increases the power 
peaks, and 
also hinders the suppression of ${\mathcal{E}}_-$ by friction; longer process times decrease power peaks, and typically ${\mathcal{E}}$ too with the exception mentioned of an ideal setting, $\gamma=0$, $\eta=1$, for which ${\mathcal{E}}=0$ for any time $t_f$.  
The contour plots of ${\mathcal{E}}$ for $\eta=\pm1$ are  quite different, see Fig. \ref{energy}, with $\mathcal{E}$ independent of $M$ if $\eta=1$, and nearly independent of $\gamma$, for weak friction, 
if $\eta=-1$. 
  
The feasibility of a given STA will not only depend on the  additive  energy consumption ${\mathcal{E}}$, but on the possibility to deliver the instantaneous power peaks, which increase
with diminishing process times. STA can be designed to lower the peak in ${\mathcal{P}}$, as it was done for $P$ in \cite{yangyang}.
The mean value theorem provides bounds for the peak of  ${\mathcal{P}}$
in different regimes 
dominated by one of the terms  in Eq. (\ref{Preal}): ${\mathcal{P}}\ge Md^2/t_f^3$
for a regime dominated by the trolley frictionless dynamics ($M$ term), whereas ${\mathcal{P}}\ge \gamma d^2/t_f^2$ for a friction dominated one. Finally, peak bounds  for $M=\gamma=0$ scale as $md^2/t_f^3$ at long process times and as $4md^2/(\omega^2 t_f^5)$ at short times.
(The bounds at short times are only meaningful for a pure harmonic oscillator since the pendulum will abandon the small-oscillations regime,
and we have assumed $\sqrt{2E_0/m}/\omega<<d$.) Minimal times for a given maximal power can be read directly from the bounds.      
%
%
%
%
%
\section{Protocol for minimal energy consumption\label{minimal}}
We will use the degeneracy of the STA to design the protocol that minimizes energy consumption, 
combining inverse engineering STA with optimal control theory \cite{expOCT, OCTtransport}. 
In this section we assume that the harmonic model holds. 

%
%
It is convenient to use the horizontal position of the load in the lab frame, $X\equiv q+x$, which obeys  
the Newton equation
\beq
\label{dynX}
\ddot X+\omega^2(X-x)=0. 
\eeq
Similarly to the difference between a general $q$ and a particular trajectory $\alpha$ in the previous section, we distinguish   
a particular trajectory $\xi$ that satisfies Eq. (\ref{dynX}) and the 
boundary conditions $\xi(0)=0, \xi(t_f)=d$ and $\dot\xi(t_b)=\ddot\xi(t_b)=0$, with $t_b=0, t_f$.  
To follow the usual conventions in optimal control theory, we use a new notation,
\beqa
y_1 = \xi,~ y_2 = \dot{\xi}, ~u(t) = x,
\eeqa
where $y_1, y_2$ are the components of a ``state vector'' $\bf{y}$, and the trolley position $u (t)$ is considered
as the (scalar) control function. With this notation
Eq. (\ref{dynX}) for $\xi$ becomes
\beqa
\label{system1}
\dot{y}_1  &=&  y_2,
\\
\label{system2}
\dot{y}_2 &=& - \omega^2( y_1- u).
\eeqa
The optimal control problem is to
find $|u (t)| \leq \delta$ for some fixed bound $\delta$,
with $u(0)=0$ and $u(t_f) =d$,  such that the system starts
at $\{y_1(0)=0, y_2 (0)= 0\}$,
ends up at $\{y_1(t_f)=d, y_2 (t_f)= 0\}$, and minimizes a cost function $J$.

In order to match the boundary conditions at the initial and final times, the optimal control obtained may be complemented by appropriate jumps.
We use Pontryagin's maximum principle, which provides necessary conditions for optimality \cite{LSP}.
Generally, to minimize the cost function
\beq
J (u)= \int^{t_f}_0 g [\textbf{y}(t), u] dt,
\eeq
the maximum principle states that for the dynamical system
$
\dot{\textbf{y}} = \textbf{f} [\textbf{y}(t),u],
$
the coordinates of the extremal vector $\textbf{y} (t)$ and of the corresponding adjoint state $\textbf{k} (t)$ formed by Lagrange multipliers, $k_1$, $k_2$, fulfill the Hamilton's equations
for a control Hamiltonian $H_c$,
\beqa
\label{H-1}
\dot{\textbf{y}} = \frac{\partial H_c}{\partial \textbf{k}},
\\
\label{H-2}
\dot{\textbf{k}} = - \frac{\partial H_c}{\partial \textbf{y}},
\eeqa
where $H_c$ is defined as
\beq
H_c [\textbf{k}(t),\textbf{y}(t),u] = k_0 g [\textbf{y}(t), u]+ \textbf{k}^{T}\cdot \textbf{f} [\textbf{y}(t),u].
\eeq
The superscript ``$T$" used here denotes the transpose of a vector,  and $k_0< 0$ can be chosen 
for convenience since it amounts to multiply the cost function by a constant. The (augmented) vector with components $(k_0, k_1, k_2)$ is nonzero and continuous. Note that 
the Lagrange multiplier $k_0$ is a constant, however, $k_1$ and $k_2$ are time dependent since the equations of motion (\ref{system1}) and (\ref{system2}) must be satisfied at all times.
For almost all $0 \leq t \leq t_f$ the function $H_c [\textbf{k}(t),\textbf{y}(t),u]$ attains its maximum at $u=u^{*}$, and
$H_c [\textbf{k}(t),\textbf{y}(t),u^{*}] =c$, where $c$ is constant.
%
%
%
%
%
%
%
%
%
%
Assuming that the integrals of two of the terms of the total power (\ref{Preal}) 
depending on $M$ and $m$ vanish (this is explicitly confirmed later)  we shall only consider the term $\gamma\dot x^2$, 
so the cost function is 
\beq
\label{Jp}
J_{\mathcal{P}}=\int_0^{t_f}\dot x^2 dt=\int_0^{t_f}\dot u^2 dt,
\eeq
for an ``unbounded problem'' (i.e., without restrictions on the possible values of the control), and an ideal $(\eta=1)$-type of process with 
perfect regenerative braking. 
The control Hamiltonian is 
\beq
H_c(k_1,k_2,y_1,y_2,u)=k_0\dot u^2+k_1y_2-k_2\omega^2(y_1-u),
\eeq
that sets the costate equations
\beqa
\dot k_1&=&\omega^2 k_2,
\nonumber \\
\dot k_2&=&-k_1. 
\eeqa
The solution to this set of equations is
\beqa
\label{k2}
k_1(t)&=&c_1\cos(\omega t)+\omega c_2\sin(\omega t),
\nonumber \\
k_2(t)&=&c_2\cos(\omega t)-\frac{c_1}{\omega}\sin(\omega t), 
\eeqa
where $c_1$ and $c_2$ are arbitrary constants. 
According to the Pontryagin's maximum principle, the time-optimal control $u(t)$ maximizes the control Hamiltonian $H_c$. 
By the Euler-Lagrange equation this is done when $u$ satisfies $k_2\omega^2=2k_0\ddot u$. Using Eq. (\ref{k2})
we find
%
%
%
%
%
%
%
\begin{figure}[t]
\begin{center}
\includegraphics[width=4.25cm]{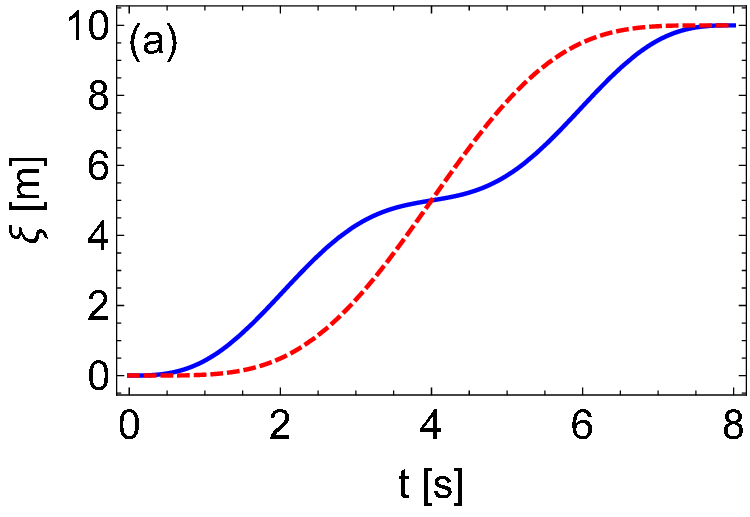}
\includegraphics[width=4.25cm]{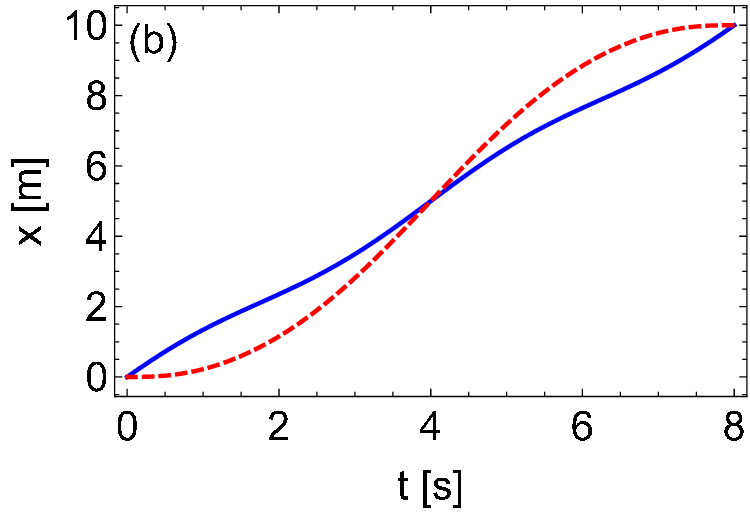}
\caption{(Color online) (a) Designed function $\xi=\alpha+x$ as a function of time. Polynomial interpolation used in Sec. \ref{secII} (red-dashed) and optimal
solution to minimize the energy consumption (blue-solid). (b) Respectively trolley displacement $x$ as a function of time. Parameter values: 
$t_f=8$ s, $l=5$ m, $d=10$ m, $g=9.8$ m/$s^2$, and $k_0=-1$. 
}
\label{xi_oct} 
\end{center}
\end{figure}
%
%
%
%
%
%
%
%
%
%
%
\beq
u(t)=x(t)=c_3+tc_4-\frac{c_2}{2k_0}\cos(\omega t)+\frac{c_1}{2k_0\omega}\sin(\omega t),
\eeq
with $c_3$ and $c_4$ also arbitrary constants. Finally solving the differential Eq. (\ref{dynX}) the optimal $\xi(t)$ is found.
The constants are fixed by imposing
the boundary conditions on $\xi$. In Fig. \ref{xi_oct}a we plot the optimal function $\xi$ and the one deduced from Sec. \ref{secII} 
with a polynomial $\alpha$. The optimal trolley displacement $x_{op}(t)$, Fig. \ref{xi_oct}b, satisfies $x_{op}(0)=0$ and $x_{op}(t_f)=d$,  
\beqa
x_{op}(t)&=&[d(-2+\omega^2t_ft+2\cos(\omega t)-2\cos(\omega(t-t_f))
\nonumber\\
&+&2\bar{c}+\omega t\bar{s})] / [-4+t_f^2\omega^2+4\bar{c}+\omega t_f\bar{s}],
\label{xop}
\eeqa
with $\bar c=\cos(\omega t_f)$ and $\bar s=\sin(\omega t_f)$. However, $\dot x_{op}(0^+)=\dot x_{op}(t_f^-)\neq 0$, $\ddot x_{op}(0^+)=-\ddot x_{op}(t_f^-)\neq 0$,
and instantaneous jumps are required to satisfy the boundary conditions $\dot x(0^-)=\dot x(t_f^+)=\ddot x(0^-)=\ddot x(t_f^+)=0$ where plus (or minus) represents an approach from the right (or left). 
The trajectory (\ref{xop}) must be limited to the domain $0<t<t_f$, and be complemented by
$x_{op}=0$, for $t<0$, and $x_{op}=d$ for $t>t_f$.   
$\dot x$ is discontinuous at $t=0$ jumping from $0$ to $x_{op}(0^+)$. Similarly at $t_f$, $\dot x$ jumps from $\dot x(t_f^-)$ to $0$. 
The acceleration thus includes Dirac-delta impulses \cite{Li-dirac,yangyang},   
%
\beq\label{x2op}
\ddot x_{op}=\left\{
\begin{array}{ll}
0, & t\leq 0^-
\\
\dot x_{op}(0^+)\delta(t),& 0^-<t<0^+
\\
\ddot x_{op}(t),& 0^+\leq t \leq t_f^-
\\
-\dot x_{op}(t_f^-)\delta(t-t_f),& t_f^-<t<t_f^+
\\
0, & t_f^+\leq t
\end{array}
\right.,
\eeq
where $\dot x_{op}$ and $\ddot x_{op}$ represent the first and second time derivative of Eq. (\ref{xop}). 
This  implies that $q$, $X$ and $\dot{X}$ are continuous at the edges.  
The protocol, including the jumps, is indeed a shortcut, as the  mechanical energy of the load, $E(t)=m(\dot x+\dot q)^2/2+m\omega^2q^2/2$, 
is equal at initial ($0^-$) and final ($t_f^+$) times.  
This can be seen from the vanishing of the integral 
\beq
\label{rterms}
  \int_{0^-}^{t_f^+}q\dot x_{op}\ dt=0,  
\eeq
which does not get any contribution at the  edges, $E(0^-)=E(0^+)=E(t_f^-)=E(t_f^+)$.  
Comparing explicitly load mechanical energies immediately before and after the boundary times this is consistent with the following 
jumps in $\dot{q}$, 
\beqa
\dot{q}(0^+)&=&\dot{q}(0^{-})-\dot{x}(0^+),
\\
\dot{q}(t_f^+)&=&\dot{q}(t_f^-)+\dot{x}(0^-). 
\eeqa
The total mechanical energy, 
\beq
E_{tot}(t)=E(t)+\frac{1}{2}M\dot x^2,
\eeq
is also equal at initial and final times since 
the trolley begins and ends at rest, 
%
%
\beq
\int_{0^-}^{t_f^+}\ddot x_{op}\dot x_{op}\ dt=0.
\eeq
In more detail, the integral vanishes in the interior domain, from $0^+$ to $t_f^-$, since $\dot x_{op}(0^+)=\dot x_{op}(t_f^-)$, 
and the jumps due to initial and final delta impulses compensate,   
$\int_{0^-}^{0^+}M\ddot x_{op}\dot x_{op}dt=M\dot x_{op}^2(0^+)/2$, and   
$\int_{t_f^-}^{t_f^+}M\ddot x_{op}\dot x_{op}dt=-M\dot x_{op}^2(t_f^-)/2$. 
%
Moreover, since the singularity of $\dot x_{op}$ at the boundaries corresponds to a finite
jump, 
\beqa
\int_{0^-}^{0^+}\dot x_{op}^2dt&=&0,
\nonumber\\
\int_{t_f^-}^{t_f^+}\dot x_{op}^2dt&=&0, 
\eeqa
%
the Dirac impulses do not contribute to the energy dissipated by friction. 
%
%
Using the expression (\ref{xop}) for the optimal trajectory we find the explicit expression of the minimal energy consumption.  
This sets a bound for any other process, 
\beq
\label{boundOCT}
\mathcal{E}\geq \frac{\gamma d^2}{t_f+\frac{4[-1+\cos(\omega t_f)]}{\omega[\omega t_f+\sin(\omega t_f)]}},
\eeq
tighter than Eq. (\ref{lower}),  $\mathcal{E} \geq\gamma d^2/t_f$. 
At large times, compared to the oscillation period, they coincide. 
Indeed $\gamma d^2/t_f$ agrees with Landauer's prediction on the energy dissipation proportional to the ``velocity of the process'' 
when there is no information loss \cite{Landauer}. However, whereas he emphasized that the dissipation can be made arbitrarily small for sufficiently long times, STA are by construction intended as fast processes where the dissipation due to friction does not vanish.   
A second difference with Landauer's discussion is that 
at short times, the dependence in Eq. (\ref{boundOCT}) changes to  
\beq
\mathcal{E}\gtrsim\frac{720d^2}{\omega^4t_f^5}, 
\eeq
with the caveat that this result indeed requires  harmonic oscillator dynamics. 

Note that the discontinuities in the derivatives of $x_{op}(t)$ imply infinite power peaks, but the energy consumed by the engine controlling the motion of the trolley, which is equal to the dissipated energy since the initial and final mechanical energies are equal,  is finite.
The ability to approach this ideal scenario of infinite power peaks will depend on the characteristics of the engine but, in any case, the bound (\ref{boundOCT}) sets the minimum energy
required to produce a STA protocol for a given transport time $t_f$.
\section{Discussion}
We have worked out an explicit model to analyze the energy consumption in shortcuts to adiabaticity.  
The model helps to point out a number of fundamental aspects, such as the importance of considering the control system
together with the primary system.  
In our model the power for the primary system and the total power only agree in a rather unrealistic scenario, namely, a control system with zero mass and no friction. The small mass limit of the control system is not only unrealistic but also undesirable, as it would make   
the total power and the external actuating control force depend on the specific dynamics (i.e., the initial boundary conditions) of the primary system.  This is against the spirit of useful shortcuts, intended to take systems from initial to final Hamiltonian 
configurations without final excitation, irrespective of the initial conditions. Control systems for microscopic primary systems will typically involve macroscopic masses, currents, or classical fields, so the need to consider the control system to examine energy costs will be prevalent.

The model also provides an ideal testbed to realize that different types of braking affect the results dramatically;  it illustrates that the stability of a given control protocol with respect to the primary system dynamics implies an energy cost and higher power peaks; and it underlines the importance of both integrated and local-in-time quantities to determine the feasibility of shortcuts.  

The current analysis may be extended to further classical, quantum, or hybrid  systems.    
In particular a quantum ``load'' represented by a particle in a harmonic trap could be driven by exactly the same STA protocols devised here,
since $I$ and $H$ have the same form as in our model.    
Close to the current model is the  transport of ions or neutral atoms for which different experiments have been performed  or are planned \cite{Couvert,trans1,trans2}. For  the transport of ultracold atoms in \cite{Couvert}, the trap was formed by optical tweezers, moved by displacing 
a lens mounted on a motorized translation stage. This setting  realizes the stabilizing $M>m$ limit, a typical scenario with microscopic loads.  
Similarly, Zenesini et al.  moved an optical lattice by displacing the mirror mounted on piezoelectric actuators \cite{Oliver}.  
For ion transport in linear, multielectrode Paul traps,  
the cost will involve assessing the energy consumed by the microchip controlling the effective moving trap 
by means of time-varying electrode potentials. The stabilization of the total power will depend on the  
macroscopic charges in the electrodes to change the voltages being much larger than the ion charge.     

While the results have been so far for a harmonic potential, 
deviations from the harmonic approximation could be 
taken into account following  \cite{NJP}.  
We may also consider 
%
\begin{figure}[h]
\begin{center}
\includegraphics[width=0.95 \linewidth]{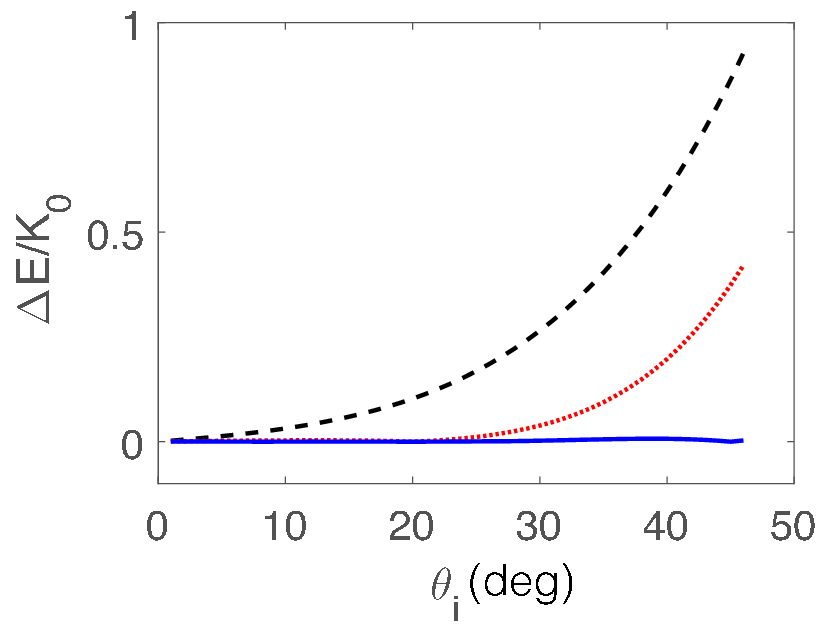}
\caption{\label{energySUP} Energy excitation of the load versus initial angle (with load initially at rest) in an
 inversely engineered transport process with $d=10$ m, $t_f=10$ s, $l=5$ m and $m=10$ kg, using additional free parameters in the ansatz for $\alpha(t)$. 
The scaling factor is the kinetic energy for a constant velocity process, $K_0=md^2/(2t_f^2)$. 
Black dashed line: process without additional parameters; red dotted line: 
one parameter is added 
to minimize excitation in $\theta_i=20^{\circ}$ ($b_8=-3513.3$); solid blue line:  minimization for the excitation in $\theta_i=20^{\circ}$ and $\theta_i=45^{\circ}$ using two free parameters ($b_8=-13862$ and $b_9=2941.5$).
}
\end{center}
\end{figure}
%
%
%
%
%
initial angles of the load $\theta_i$ beyond the small oscillations regime and redesign the protocol for the trolley motion $x(t)$ to minimize the difference between initial and final mechanical energies of the load ($\Delta E = |E_f - E_0|$). 
This requires a higher order polynomial functions $\alpha(t)=\sum_{j=0}^{7+n} b_j t^j$ to minimize the energy difference for one or more ($n$) initial angles $\theta_i$ with the extra parameters.   
The number of free parameters $n$ is set by the number of initial angles used to minimize the excitation, and the rest of coefficients in $\alpha(t)$ are fixed by the boundary conditions as Sec. \ref{secII}. In Fig. \ref{energySUP} we plot the excitation energy for
processes with one and two free parameters, and for the process in Sec. \ref{secII} ($n=0$). The figure demonstrates clearly that STA beyond the small oscillation regime are indeed possible. This implies zero or negligible energy consumption under ideal conditions
(no friction, $\gamma=0$, and regenerative braking, $\eta=1$). 

For a general system, beyond transport systems,  regardless of the specific dynamics involved,    
friction, the combination of positive and negative power domains, and the independence of the external forces with respect to 
the primary system dynamics will be ubiquitous in STA implementations, 
and thus essential elements to evaluate actual energy consumptions.   
Whereas for slow processes, the energy dissipated by friction can be made negligible (a standard assumption 
for infinite-time processes), even if the friction coefficient is not zero, STA are by definition fast processes, so to neglect energy dissipation in STA the stronger assumption of zero friction coefficients
is necessary.  
Again, the fast nature of STA protocols implies large positive and negative powers which enhances the importance of braking.      
Braking mechanisms determine the cost of the energy integrated in negative power segments, and if it can indeed be reused.  
In typical scenarios this is not the case, i.e., $\eta\ne 1$, so that negative power segments consume energy (the extreme case is $\eta=-1$) or, if they do not 
consume energy ($\eta=0$), they do not compensate for the consumption in positive segments.     
For the realistic expectation that $\gamma \ne0$ and $\eta\ne 1$, shorter process times imply higher power peaks and 
an increased energy consumption. 
Note that even in the highly idealized limit $\gamma=0$, $\eta=1$, with zero global cost, (with respect to final adiabatic energy minus initial energy), shortening the time also implies higher power peaks, which become a limiting factor that cannot be ignored to determine the feasibility of a shortcut.

%
We thank D. Gu\'ery-Odelin and A. Levy
for discussions.
We acknowledge  
funding by the Basque Government (Grant No. IT986-16), MINECO/FEDER,UE (Grants FIS2015-67161-P and FIS2015-70856-P), and QUITEMAD+CM S2013-ICE2801.
%
%
%

\end{document}